\documentclass[a4paper]{jpconf}
\usepackage{graphicx}
\usepackage{lineno}
\begin{document}
\title{Large-Scale Distribution of Arrival Directions of Cosmic Rays Detected at the Pierre Auger Observatory Above $10~$PeV}

\author{Olivier Deligny$^1$, for the Pierre Auger Collaboration$^2$}

\address{$^1$ IPN Orsay, CNRS/IN2P3 \& Universit\'e Paris Sud, 15 rue Clemenceau, 91406 Orsay CEDEX, France\\
$^2$ Observat\'orio Pierre Auger, Av. San Mart\'in Norte 306, 5613 Malarg\"ue, Mendoza, Argentina; {\tt www.auger.org};
full author list: {\tt http://www.auger.org.ar/archive/authors\_2013\_07.html}}

\ead{deligny@ipno.in2p3.fr}

\begin{abstract}
Searches for large-scale anisotropies in the distribution of arrival directions of cosmic rays detected above 
$\simeq 10$~PeV at the Pierre Auger Observatory are presented. Although no significant deviation from isotropy
is revealed at present, some of the measurements suggest that future data will provide hints for large-scale
anisotropies over a wide energy range. Those anisotropies would have amplitudes which are too small to be significantly 
observed within the current statistics. Assuming that the cosmic ray anisotropy is dominated by dipole and quadrupole
moments in the EeV-energy range, some consequences of the present upper limits on their amplitudes are presented.
\end{abstract}


\section{Introduction}
\label{sec:intro}

The large-scale distribution of arrival directions of cosmic rays is an important observable in attempts to understand 
their origin. This is because this observable is closely connected to both their source distribution and their propagation. 
Due to scattering in magnetic fields, the anisotropy imprinted in the arrival directions is mainly expected at large 
scales up to the highest energies. For energies above a few PeV, the decrease of the intensity with energy makes it 
challenging to collect the necessary statistics required to reveal amplitudes down to $10^{-3}-10^{-2}$. 

The surface detector arrays operating at the Pierre Auger Observatory provide high-quality data to search for 
large-scale anisotropies from $\simeq$10~PeV up to the highest energies. The different analyses performed by the 
Pierre Auger Collaboration on data taken from 2004 until the end of 2012 are summarised in this paper. Comprehensive 
reports describing all the details of the analyses can be found in~\cite{AugerAPP2011,AugerApJL2013,AugerApJS2012,
Sidelnik2013,Menezes2013,Deligny2013,AugerTA2014}. An emphasis is given here to hints that may be 
indicative of anisotropies whose amplitudes are still too small to stand out significantly above the background noise 
which is intrinsic to the finite statistics accumulated so far, and to the implications on the origin of EeV-cosmic rays that 
can be inferred from the upper limits obtained in this energy range.

The 1500~m surface detector array of the Pierre Auger Observatory is composed of 1600 water-Cherenkov detectors 
arranged in a triangular grid with spacing of 1500~m, covering a total area of 3000 km$^2$. This array is fully efficient 
at energies above 3~EeV. 49 additional detectors with 750~m spacing have been nested within the 1500~m array to 
cover an area of 25 km$^2$ with full efficiency above 300~PeV. This 750~m surface detector array allows complementary
anisotropy searches at much lower energies, down to $\simeq$ 10~PeV.

\section{Harmonic Analyses in Right Ascension Above $10~$PeV}
\label{sec:harmonic}

\subsection{Event Rate}
\label{sec:hexagons}

To search for large-scale anisotropies, it is mandatory to account for the modulation of the event rate imprinted by variations 
of experimental origin. High-quality events used here are selected on the condition that the elemental cell of the event (that is, 
all six neighbors of the water-Cherenkov detector with the highest signal) was active at the time the event was
recorded~\cite{AugerNIM2010}. Based on this fiducial cut, the total geometric exposure to the end of 2012 was 31645 km$^2~$sr~yr for the 1500~m surface detector array and 79 km$^2~$sr~yr for the 750~m one. Since the number of
elemental cells $n_\mathrm{cell}(t)$ is accurately monitored every second, the expected number of events for a uniform flux 
can be inferred at the corresponding accuracy as a function of time. Denoting by $\alpha^0$ the local sidereal time and by
$T_{\mathrm{sid}}$ the sidereal period, the total number of elemental cells $N_{\mathrm{cell}}(\alpha^0)$ and its 
associated relative variations $\Delta N_{\mathrm{cell}}(\alpha^0)$ can then be obtained as:
\begin{equation}
\label{Ncell}
N_{\mathrm{cell}}(\alpha^0)=\sum_{j}n_{\mathrm{cell}}(\alpha^0+jT_{\mathrm{sid}}), \hspace{1cm} \Delta N_{\mathrm{cell}}(\alpha^0)=\frac{N_{\mathrm{cell}}(\alpha^0)}{\left<N_{\mathrm{cell}}\right>_{\alpha^0}},
\end{equation}
with $\left<N_{\mathrm{cell}}\right>_{\alpha^0}=1/T_{\mathrm{sid}}\int_0^{T_{\mathrm{sid}}}\mathrm{d}\alpha^0N_{\mathrm{cell}}(\alpha^0)$.
This latter quantity is the relevant input to account for the changes in time of the observatory when estimating the anisotropy 
parameters.

\subsection{Event Rate and Weather Effects}
\label{sec:weather}

The energy estimator of the showers collected at the surface detector array of the Pierre Auger Observatory is provided 
by the signal size measured at 1000~m from the shower core, $S(1000)$. Since the development of extensive air 
showers depends on the atmospheric pressure $P$ and air density $\rho$, the corresponding $S(1000)$ for any fixed 
energy is sensitive to variations in pressure and air density. Systematic variations with time of $S(1000)$ induce variations 
of the event rate that may distort the real dependence of the cosmic ray intensity with right ascension. To cope with this
experimental effect, the observed shower size $S(1000)$, measured at the actual density $\rho$ and pressure $P$, is 
converted to the one that would have been measured at reference values $\rho_0$ and $P_0$~\cite{AugerAPP2009}. 
Applying these corrections to the energy assignments of showers allows the cancellation of spurious variations of the 
event rate in right ascension, whose typical amplitudes amount to a few per thousand when considering data sets collected 
over full years~\cite{AugerAPP2011}.

\subsection{Different Analyses}
\label{sec:differentanalyses}

To account for the non-uniform exposure in the sky, the inverse of $\Delta N_{\mathrm{cell}}(\alpha^0)$ can be used
to weight the contribution of each event entering in the calculation of the Fourier first-harmonic 
coefficients~\cite{MollerachJCAP2005}:
\begin{equation}
a=\sum_{i=1}^N \frac{\cos{\alpha_i}}{\Delta N_{\mathrm{cell}}(\alpha^0_i)}, ~~~~b=\sum_{i=1}^N \frac{\sin{\alpha_i}}{\Delta N_{\mathrm{cell}}(\alpha^0_i)}.
\end{equation}
The sum runs over the number $N$ of events in the energy range considered, and the normalisation reads as 
$\mathcal{N}=\sum_{i=1}^N 1/\Delta N_{\mathrm{cell}}(\alpha^0_i)$. The amplitude and phase are then calculated
as $r=\sqrt{a^2+b^2}$ and $\phi=\arctan{(b/a)}$. For isotropy, and for weight values close to 1, these random variables 
follow a Rayleigh and a uniform distributions respectively~\cite{Linsley1975}. 

This procedure has been shown to provide accurate estimates for data collected at the 1500~m surface detector array 
above 1~EeV~\cite{AugerAPP2011}. Below 1~EeV, in addition to the energy estimates, weather effects can also affect the
detection efficiency as a function of time and thus imprint fake anisotropies in the arrival directions. To circumvent this difficulty,
the estimation of the anisotropy parameters is then carried out using the East-West method~\cite{Bonino2011}, which is based
on the fact that the instantaneous exposure for Eastward and Westward events is the same. Hence, the difference between 
the event counting rate measured from the East sector and the West sector can provide unbiased estimates of anisotropy
parameters without applying any correction, although at the price of reducing the sensitivity to the first harmonic modulation.
Note also that, in the case of the 750~m surface detector array, results are presented only in terms of the East-West method.

\subsection{Measurements of First Harmonic Amplitudes}
\label{sec:amplitudes2d}

The Rayleigh amplitude $r$ measured by any observatory can be used to reveal anisotropies projected on the Earth's equatorial
plane. In the case of an underlying pure dipole, the relationship between $r$ and the projection of the dipole on the Earth's
equatorial plane, $d_\perp$, depends on the latitude of the observatory and on the range of zenith angles considered:
$d_\perp\simeq r/\left<\cos\delta\right>$. $d_\perp$ is thus the physical quantity of interest to compare the results of different
experiments and the pure dipole predictions.

\begin{figure}[!ht]
   \centering
    \includegraphics[width=10cm]{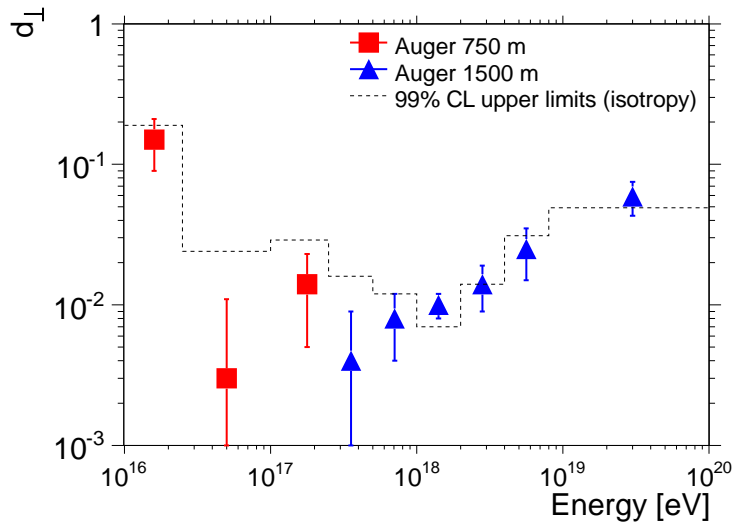}
    \caption{First harmonic amplitude of the equatorial dipole anisotropy as a function of energy. The dashed lines are the 99\%
    confidence level upper values of the amplitude that could result from fluctuations of an isotropic distribution.
    From~\cite{Sidelnik2013}.}
   \label{fig:amplitudes}
\end{figure}

The amplitudes $d_\perp$ obtained are shown in figure~\ref{fig:amplitudes}. The dashed line stands for the upper values of 
the amplitude which may arise from fluctuations in an isotropic distribution at 99\% confidence level. Note that in the energy
ranges 1-2 and 2-4 EeV the measured amplitudes have a probability to arise by chance from an isotropic distribution of about
0.03\% and 0.9\%, while above 8 EeV the measured amplitude has chance probability of only 0.1\%. Since several energy bins
were searched, these numbers do not represent absolute probabilities. They constitute however interesting hints for 
large-scale anisotropies that will be important to further scrutinise with increased statistics.

\subsection{Notes on the Measurements of First Harmonic Phases}
\label{sec:phases2d}

It has been noted in the past that, for an underlying anisotropy, the detection of the genuine phase is expected to occur 
earlier than the detection of a significant amplitude~\cite{Edge1978}. For an anisotropy evolving smoothly with
the energy, testing the consistency of independent phase measurements, such as a constancy or a smooth evolution
in adjacent energy bins, may thus be a powerful tool for detecting anisotropy.  

A likelihood ratio test can be designed to quantify whether or not a parent random distribution of arrival directions 
better reproduces the phase measurements in different energy intervals than an alternative dipolar parent 
distribution~\cite{AugerAPP2011}. The likelihood ratio $\lambda$ is built from the p.d.f. of each independent
measurement (let's suppose here $N_b$ independent bins ordered in energy, with $N=30,000$ events in each bin) 
under the hypothesis of isotropy ($p^{\rm{iso}}_\Phi(\phi)$, uniform distribution) and under the hypothesis of a dipolar
distribution ($p^{\rm{dipole}}_\Phi(\phi)$ calculated for the first time in~\cite{Linsley1975}):
\begin{equation}
\lambda=\frac{\prod_{i=1}^{i=N_b} p^{\rm{iso}}_\Phi(\phi_i)}{\prod_{i=1}^{i=N_b} p^{\rm{dipole}}_\Phi(\phi_i)}.
\end{equation}
The expected amplitude values $s$ entering into each $p^{\rm{dipole}}_\Phi(\phi_i)$ can be estimated by the expected 
mean noise given the available statistics. The statistic of the variable $-2\ln{(\lambda)}$ is then built under the hypothesis 
of isotropy by means of a large number of Monte-Carlo isotropic samples. The probability that the hypothesis of isotropy 
better reproduces the measurements compared to the alternative hypothesis is then calculated by integrating the 
normalised distribution of $-2\ln{(\lambda)}$ above the value found in the data set.

\begin{figure}[!ht]
   \centering
    \includegraphics[width=10cm]{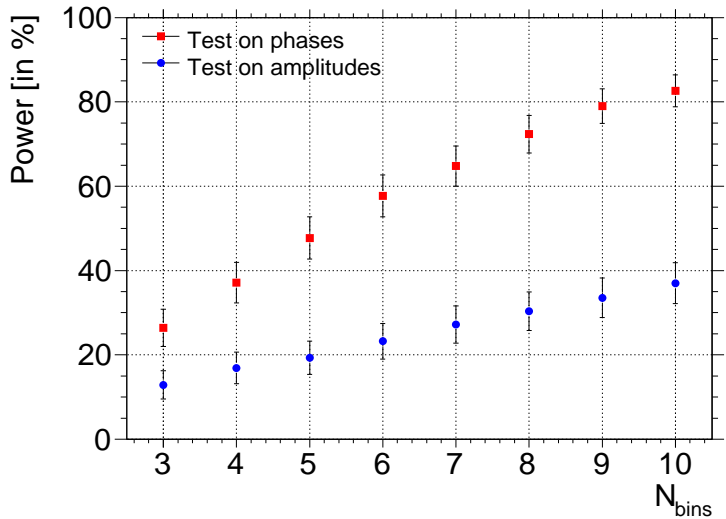}
    \caption{\small{Power of the tests on amplitudes (in blue) and on phases (in red) as a function of the number of bins
    $N_b$ entering in each test,  in the case of a genuine signal $s=1\%$ and with $N=30,000$ events in each bin.}}
   \label{fig:power}
\end{figure}

The power of this test is illustrated in figure~\ref{fig:power}, where its efficiency (shown as the red squares) to detect
a genuine anisotropy for a threshold value of 1\% is shown as a function of the number of bins $N_b$. Compared to the 
efficiency of the "$2K$" test on amplitudes (see~\cite{Edge1978} for details about the "$2K$" test), it is apparent that the
test on the consistency of the phase measurements leads to a better power by a factor greater than 2.

\begin{figure}[!ht]
  \centering
  \includegraphics[width=7.5cm]{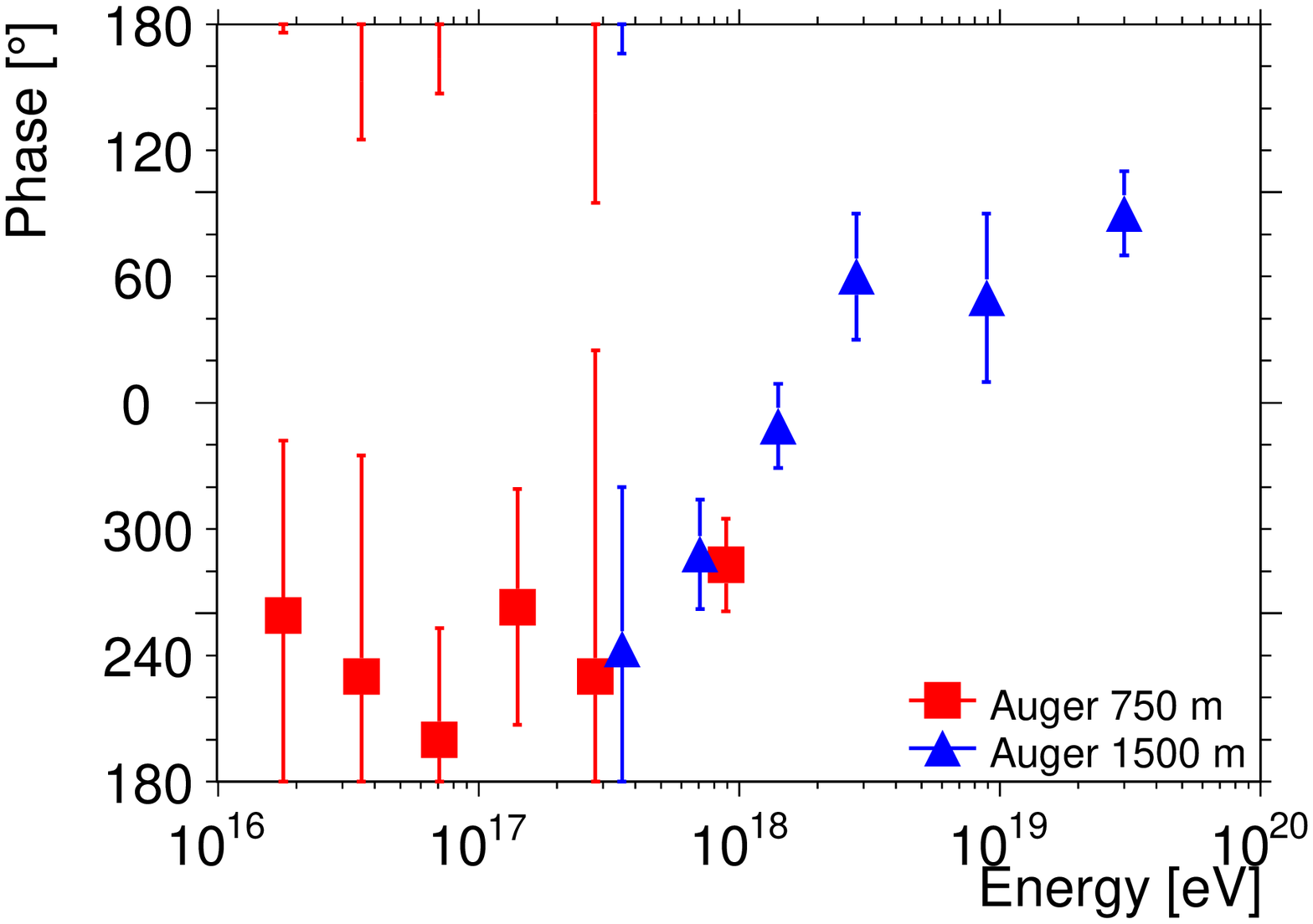}
  \includegraphics[width=7.5cm]{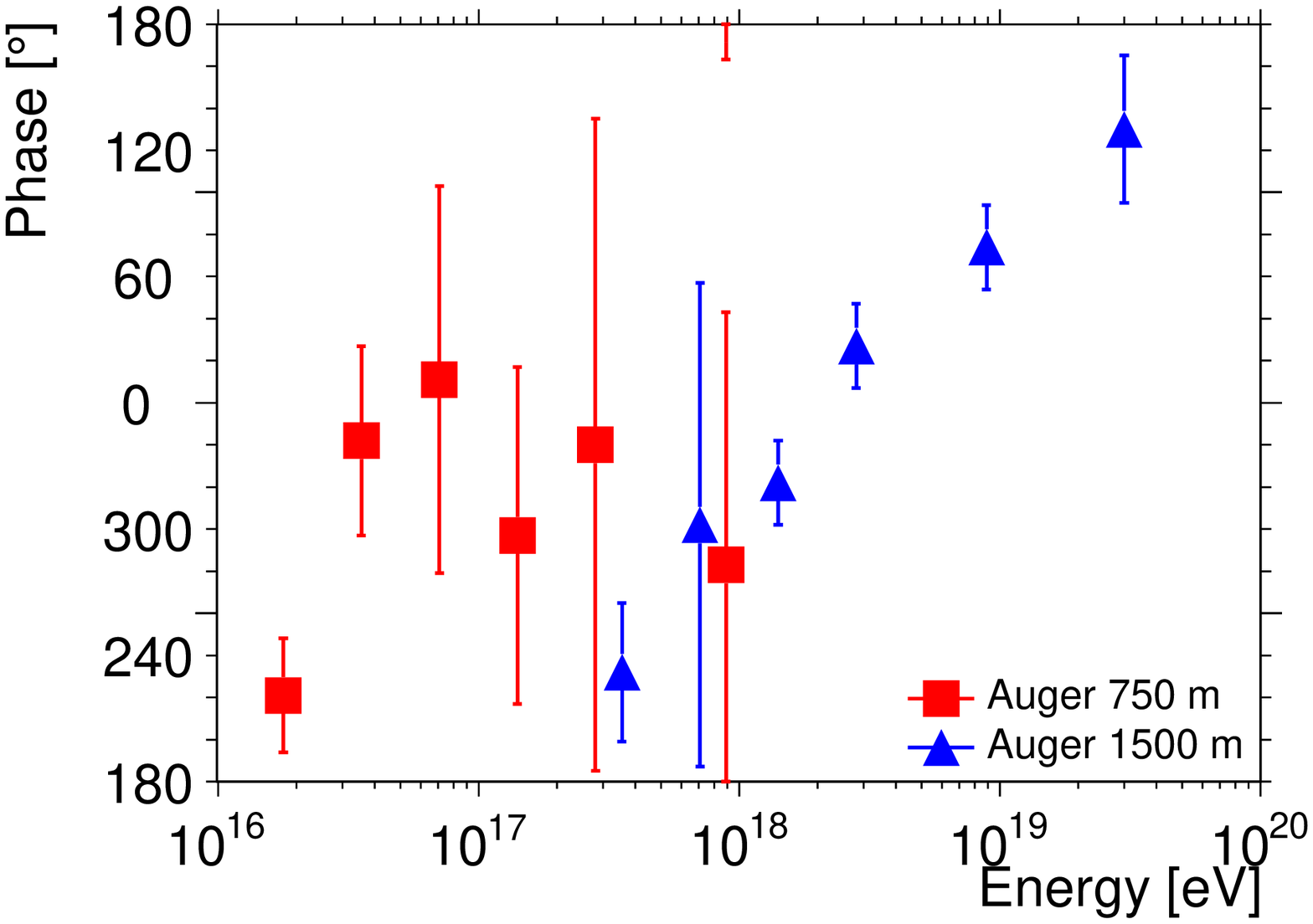}
  \caption{Phase of the first harmonic in right ascension as a function of energy. Left: Data collected from January 1 2004 to
  December 31 2010 for the 1500~m array, and from September 12 2007 to April 11 2011 for the 750~m one. Right: Data 
  from June 25 2011 (beginning of the prescription) up to December 31 2012. From~\cite{Sidelnik2013}.}
  \label{fig:phases}
\end{figure}

Driven by this property, it was pointed out in~\cite{AugerAPP2011} that the phase measurements in adjacent energy 
intervals above $\simeq 100~$PeV was suggestive of a smooth transition between $\simeq 270^\circ$ below 1~EeV 
and another phase ($\simeq 100^\circ$) above 5~EeV. This behaviour motivated the Pierre Auger Collaboration to design 
a prescription with the intention of establishing at 99\% confidence level whether this consistency in phases in adjacent 
energy intervals is real or not. The test makes use of all high-quality events above 10~PeV collected at the 750~m 
surface detector array, and all high-quality events above 250~PeV collected at the 1500~m surface detector array. 
The mid-term status of the prescription, reported in~\cite{Sidelnik2013}, is illustrated in figure~\ref{fig:phases}. The right 
panel is derived with data collected since June 25 2011 (beginning of the prescription) up to December 31 2012 only. At this 
stage, the values as derived from the analysis applied to the 750~m array are still affected by large uncertainties. On the 
other hand, the overall behavior of the points as derived from the analysis applied to the 1500~m array shows good 
agreement with the previous observations. The final result of the prescription is expected for 2015, once the required 
exposure in the prescription is reached.

\section{Spherical Harmonic Analyses Above $1~$EeV}
\label{sec:spherical_harmonic}

\subsection{Event Rate and Geomagnetic Field}
\label{sec:gmf}

To characterise the arrival directions in both right ascension and declination angles, it is essential to keep under
control the observed rate in terms of local angles. The geomagnetic field turns out to influence the shower
developments and shower size estimations at a fixed energy due to the broadening of the spatial distribution 
of particles in the direction of the Lorentz force. As the strength of the geomagnetic field component perpendicular 
to any arrival direction depends on both the zenith and azimuthal angles, the small changes of the density of 
particles at ground break the circular symmetry of the lateral spread of the particles and thus induce a dependence 
of the shower size $S(1000)$ at a fixed energy in terms of the azimuthal angle. Due to the steepness of the
energy spectrum, such an azimuthal dependence translates into azimuthal modulations of the observed cosmic 
ray event rate at a given $S(1000)$. To eliminate these effects, the observed shower size $S(1000)$ is thus 
converted to the one that would have been observed in the absence of geomagnetic field~\cite{AugerJCAP2011}. 

\subsection{Directional Exposure}
\label{sec:exposure}

The \textit{directional exposure} $\omega$ of the Observatory, providing the effective time-integrated collecting 
area for a flux from each direction of the sky in unitsof km$^2$~yr, is the key-ingredient to be determined accurately in
searching for anisotropies. The directional aperture per cell can be known in a pure geometric way as:
\begin{equation}
\label{eqn:aperture}
a_{\mathrm{cell}}^{(i)}(\theta,\varphi)=1.95~[1+\zeta^{(i)}\tan{\theta}\cos{(\varphi-\varphi_0^{(i)})}]~\cos{\theta}~\mathrm{km}^2,
\end{equation}
where $\zeta^{(i)}$ and $\varphi_0^{(i)}$ are the zenith and azimuth angles of the normal vector to each elemental 
cell. Given that the number of high-quality selected events is proportional to the number $n_{\mathrm{cell}}(t)$ of 
available cells at any time, the directional exposure is controlled by this number 
$n_{\mathrm{cell}}(t)$~\cite{AugerNIM2010}. In addition, for energies below 3~EeV, it is also controlled by the 
detection efficiency $\epsilon$ for triggering. This efficiency depends on the energy $E$, the zenith angle $\theta$, 
and the azimuth angle $\varphi$. 

To determine the detection efficiency function, an empirical approach based on the quasi-invariance of the zenith
distribution to large-scale anisotropies for zenith angles less than $\simeq 60^\circ$ and for any Observatory whose 
latitude is far from the poles of the Earth can be adopted~\cite{AugerApJS2012}. For full efficiency, the distribution
$dN/d\sin^2{\theta}$ is expected to be uniform for geometry reasons. Consequently, below full 
efficiency, deviations from a uniform behaviour in the $dN/d\sin^2{\theta}$ distribution provides 
an empirical measurement of the zenith dependence of the detection efficiency. Since this is applied to all events 
detected without distinction based on the primary mass of cosmic rays, this technique does not provide the mass 
dependence of the detection efficiency. For that reason, the anisotropy searches reported in~\cite{AugerApJS2012,
Menezes2013} and presented below pertain to the whole population of cosmic rays, whether this population consists 
of a single primary mass or a mixture of several elements.

Because the observed azimuthal distribution $dN/d\varphi$ is, in contrast, sensitive to large-scale
anisotropies~\cite{AugerApJS2012}, the same empirical approach cannot be used to determine the variations with
$\varphi$ around the mean value of $\epsilon$ previously derived for each energy and each zenith angle. The effects
inducing these variations have thus to be modelled accurately. 

The first effect is due to the influence of the geomagnetic field. Indeed, if the procedure consisting in correcting
the energies for the geomagnetic effects guarantees an unbiased event rate for full efficiency, it does not account
for the fact that under any incident angles $(\theta,\varphi)$ below full efficiency, the observed showers
triggered the surface detector array with a probability associated to their \textit{uncorrected} sizes. Above 1~EeV,
this effect is the main source of azimuthal dependence of the detection efficiency. It can be accounted for in a
straightforward way by retro-propagating the angular dependence of the geomagnetic corrections into the energy
given in the argument of the mean $\epsilon$ function determined previously. 

The second relevant effect is induced by the tilt of the array. This is because the effective separation between 
detectors for a given zenith angle depends actually on the azimuth. Since the surface detector array seen by showers
coming from the uphill direction is denser than that for those coming from the downhill one, the detection efficiency is
higher in the uphill direction. The corresponding change in detection efficiency 
$\Delta\epsilon_{\mathrm{tilt}}(\theta,\varphi,E)$ can be parameterised in a quite generic way~\cite{AugerApJS2012}. 

Overall, the final expression used to calculate the directional exposure is the following:
\begin{equation}
\label{eqn:exposure}
\omega(\delta,E)= \sum_{i=1}^{n_{\mathrm{cell}}}x^{(i)}\int_0^{24h}~\mathrm{d}\alpha^\prime\,a_{\mathrm{cell}}^{(i)}{(\theta,\varphi)}~\left[\epsilon(\theta,\varphi,E)+\Delta\epsilon_{\mathrm{tilt}}(\theta,\varphi,E)\right],
\end{equation}
where $x^{(i)}$ is the total operational time of the cell $(i)$ considered as independent of the local sidereal 
time~\footnote{Although the operational time of each cell has not been constant within each sidereal day due to 
temporary instabilities and/or failures, the resulting modulation is overall negligible once the total operational time 
obtained by summing over a large number of sidereal days is considered.}. The integration variable $\alpha^\prime$
stands for the hour angle. The transformation from local to celestial angles accounts for the slightly different
latitudes of the cells to account for the spatial extension of the surface detector array. For a wide range of declinations
between $\simeq -89^\circ$ and $\simeq 20^\circ$, the directional exposure is $\simeq 2,990~$km$^2~$yr at 1~EeV 
and $\simeq 4,180~$km$^2~$yr above 3~EeV.

\subsection{Estimates of Spherical Harmonic Coefficients}
\label{sec:alm}

Any angular distribution over the sphere $\Phi(\mathbf{n})$ can be decomposed in terms of a multipolar expansion:
\begin{equation}
\label{eqn:ylm}
\Phi(\mathbf{n})=\sum_{\ell\geq0}\sum_{m=-\ell}^{\ell}~a_{\ell m}Y_{\ell m}(\mathbf{n}),
\end{equation}
where $\mathbf{n}$ denotes a unit vector taken in equatorial coordinates. The customary recipe to extract each 
multipolar coefficient makes use of the completeness relation of spherical harmonics:
\begin{equation}
\label{eqn:alm}
a_{\ell m}=\int_{4\pi} \mathrm{d}\Omega~\Phi(\mathbf{n})Y_{\ell m}(\mathbf{n}),
\end{equation}
where the integration is over the entire sphere of directions $\mathbf{n}$. Any anisotropy fingerprint is encoded in 
the $a_{\ell m}$ spherical harmonic coefficients. Variations on an angular scale of $\Theta$ radians contribute amplitude 
in the $\ell\simeq1/\Theta$ modes. However, due the non-uniform and incomplete coverage of the sky at the Pierre Auger
Observatory, the estimated coefficients $\overline{a}_{\ell m}$ are determined in a two-step procedure. First, from any
event set with arrival directions $\{\mathbf{n_1},...,\mathbf{n_N}\}$ recorded at local sidereal times 
$\{\alpha^0_1,...,\alpha^0_N\}$, the multipolar coefficients of the angular distribution coupled to the exposure function 
are estimated through:
\begin{equation}
\label{eqn:blm-est}
\overline{b}_{\ell m}=\sum_{k=1}^{N} \frac{Y_{\ell m}(\mathbf{n}_k)}{\Delta N_{\mathrm{cell}}(\alpha^0_{k})}.
\end{equation}
The weights $1/\Delta N_{\mathrm{cell}}(\alpha^0_{k})$ are introduced to correct for the slightly non-uniform directional
exposure in right ascension. Then, assuming that the multipolar expansion of the angular distribution $\Phi(\mathbf{n})$ 
is \textit{bounded} to $\ell_{\mathrm{max}}$, the first $b_{\ell m}$ coefficients with $\ell\leq\ell_{\mathrm{max}}$ are 
related to the non-vanishing $a_{\ell m}$ through:
\begin{equation}
\label{eqn:blm-est}
\overline{b}_{\ell m}=\sum_{\ell^\prime=0}^{\ell_{\mathrm{max}}}\sum_{m^\prime=-\ell^\prime}^{\ell^\prime} [K]_{\ell m}^{\ell^\prime m^\prime} \overline{a}_{\ell^\prime m^\prime},
\end{equation}
where the matrix $K$ is entirely determined by the directional exposure:
\begin{equation}
\label{eqn:K}
[K]_{\ell m}^{\ell^\prime m^\prime}=\int_{\Delta\Omega} \mathrm{d}\Omega~\omega(\mathbf{n})~Y_{\ell m}(\mathbf{n})~Y_{\ell^\prime m^\prime}(\mathbf{n}).
\end{equation}
Inverting Eqn.~\ref{eqn:blm-est} allows a recovering of the underlying $\overline{a}_{\ell m}$, with a resolution proportional
to $([K^{-1}]_{\ell m}^{\ell m}~\overline{a}_{00})^{0.5}$~\cite{Billoir2008}. As a consequence of the incomplete 
coverage of the sky, this resolution deteriorates by a factor larger than 2 each time $\ell_{\mathrm{max}}$ is incremented 
by 1. Given the current present statistics, this prevents the recovery of each coefficient with good accuracy as soon as $\ell_{\mathrm{max}}\geq3$.

\subsection{Dipole and Quadrupole Moments}
\label{sec:dipquad}

For the reason explained just above, the angular distribution $\Phi(\mathbf{n})$ is assumed to be modulated by a dipole
and a quadrupole only. In this case, $\Phi(\mathbf{n})$ can be written in a convenient geometric way as:
\begin{equation}
\label{eqn:phi-quad}
\Phi(\mathbf{n})=\frac{\Phi_0}{4\pi}~\bigg(1+r~\mathbf{d}\cdot\mathbf{n} +\lambda_+(\mathbf{q_+}\cdot\mathbf{n})^2 +\lambda_0(\mathbf{q_0}\cdot\mathbf{n})^2 +\lambda_-(\mathbf{q_-}\cdot\mathbf{n})^2 \bigg).
\end{equation}
The dipole pattern is here fully characterised by a vector with declination $\delta_d$, right ascension $\alpha_d$,
and an amplitude $r$ corresponding to the maximal anisotropy contrast: $r=(\Phi_{\mathrm{max}}-\Phi_{\mathrm{min}})/(\Phi_{\mathrm{max}}+\Phi_{\mathrm{min}})$.
On the other hand, the quadrupolar pattern is fully characterised by the eigenvalues and the corresponding eigenvectors
of a tensor. It is convenient to define the quadrupole amplitude 
$\beta\equiv(\lambda_+-\lambda_-)/(2+\lambda_++\lambda_-)$, which provides a measure of the maximal quadrupolar 
contrast in the absence of a dipole. The quadrupole is then described in terms of two amplitudes $(\beta,\lambda_+)$ 
and three angles: $(\delta_+,\alpha_+)$ which define the orientation of $\mathbf{q_+}$ and $(\alpha_-)$ which defines 
the direction of $\mathbf{q_-}$ in the orthogonal plane to $\mathbf{q_+}$. The third eigenvector $\mathbf{q_0}$ is 
orthogonal to $\mathbf{q_+}$ and $\mathbf{q_-}$, and has an amplitude determined from the traceless constraint 
$\lambda_++\lambda_-+\lambda_0=0$. 

\begin{figure}[!ht]
  \centering
  \includegraphics[width=7.5cm]{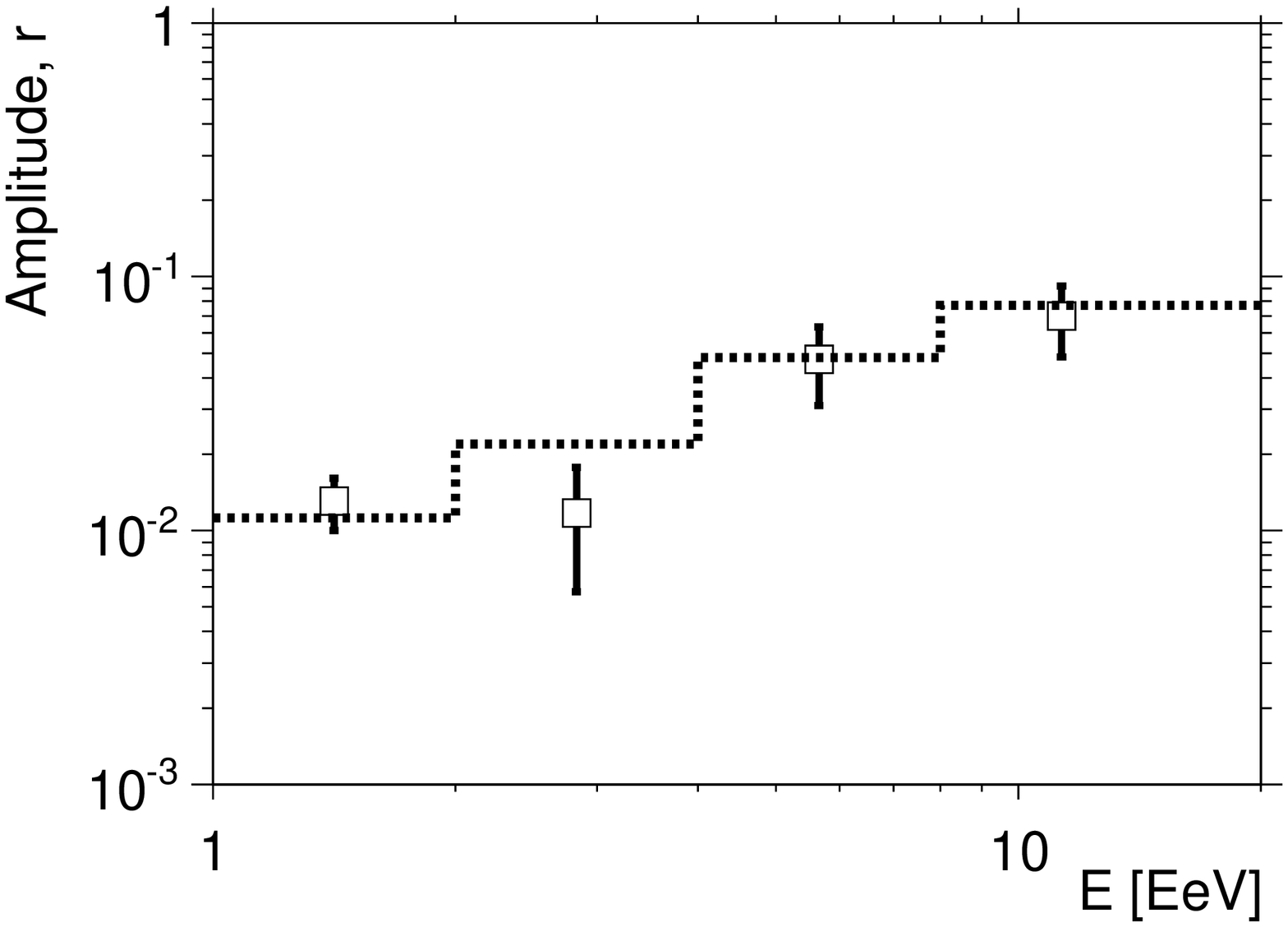}
  \includegraphics[width=7.5cm]{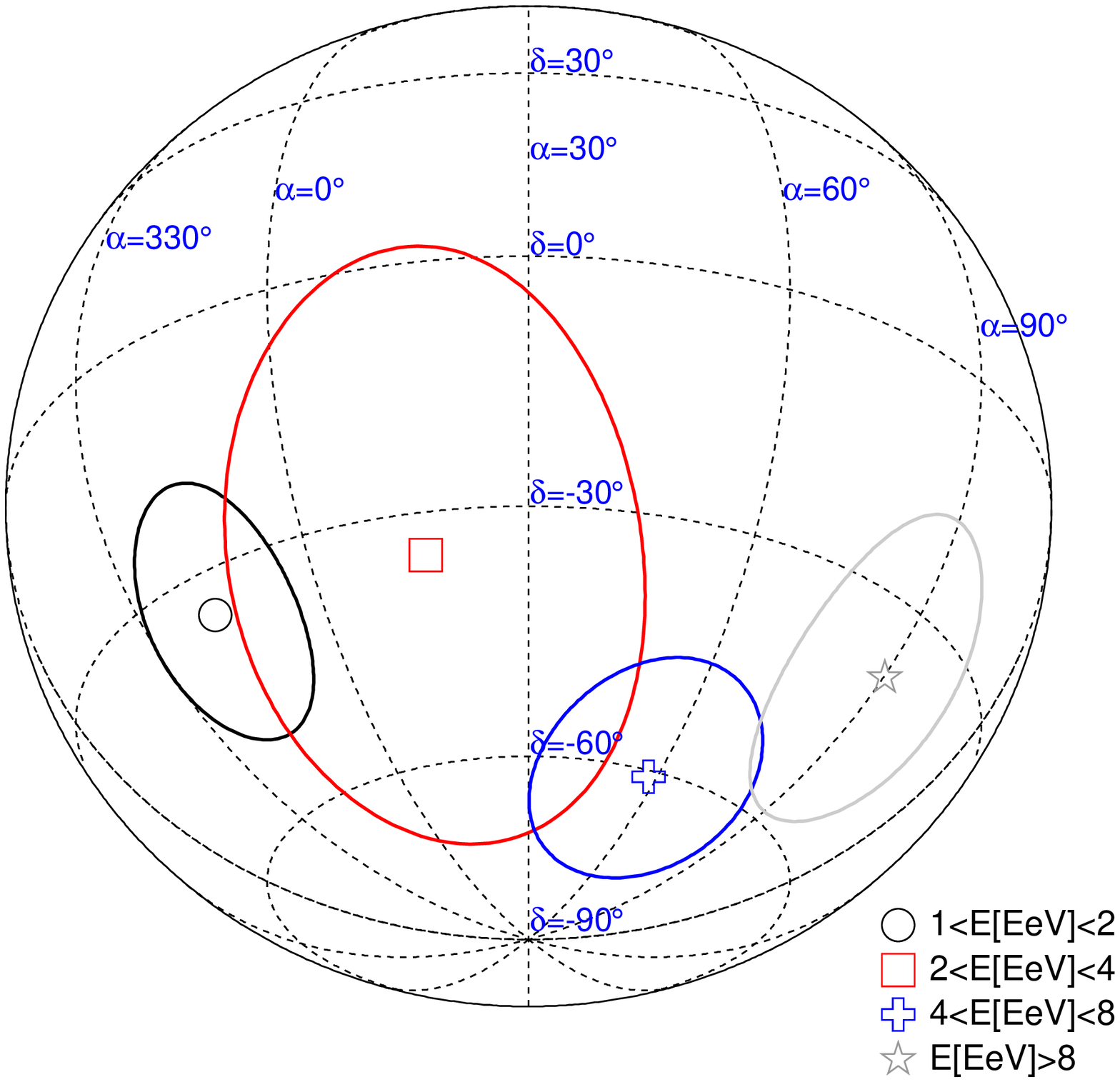}
  \caption{Left: Reconstructed amplitude of the dipole as a function of the energy. The dotted line stands for the 99\% 
  confidence level upper bounds on the amplitudes that would result from fluctuations of an isotropic distribution. Right:
  Reconstructed declination and right ascension  of the dipole with corresponding uncertainties, as a function of the energy, 
  in orthographic projection. From~\cite{Menezes2013}.}
  \label{fig:dipole}
\end{figure}

For a pure dipole ($\lambda_{\pm,0}=0$), the reconstructed amplitudes are shown in figure~\ref{fig:dipole} 
as a function of the energy. The 99\% confidence level upper bounds on the amplitudes that would result from fluctuations 
of an isotropic distribution are indicated by the dotted line. Similarly to the results presented in last section, interesting 
hints for large-scale anisotropies are observed in the first energy bin, hints that will be important to further scrutinise with 
independent data. The corresponding reconstructed directions in orthographic projection with the associated uncertainties
are also shown as a function of the energy. The same behaviour is observed for the phases as in the previous analysis.
Considering also the possibility for a non-zero quadrupole, the dipole amplitude in the first energy bin is not any longer
above the corresponding 99\% confidence level upper bound due to the lost of sensitivity when increasing the upper bound
of the multipolar expansion. 

The upper limits that can be derived on the dipole and quadrupole amplitudes are presented and discussed in the next section.



\section{Constraints on the Origin of EeV-Cosmic Rays}
\label{sec:constraints}

\begin{figure}[!ht]
  \centering
  \includegraphics[width=7.5cm]{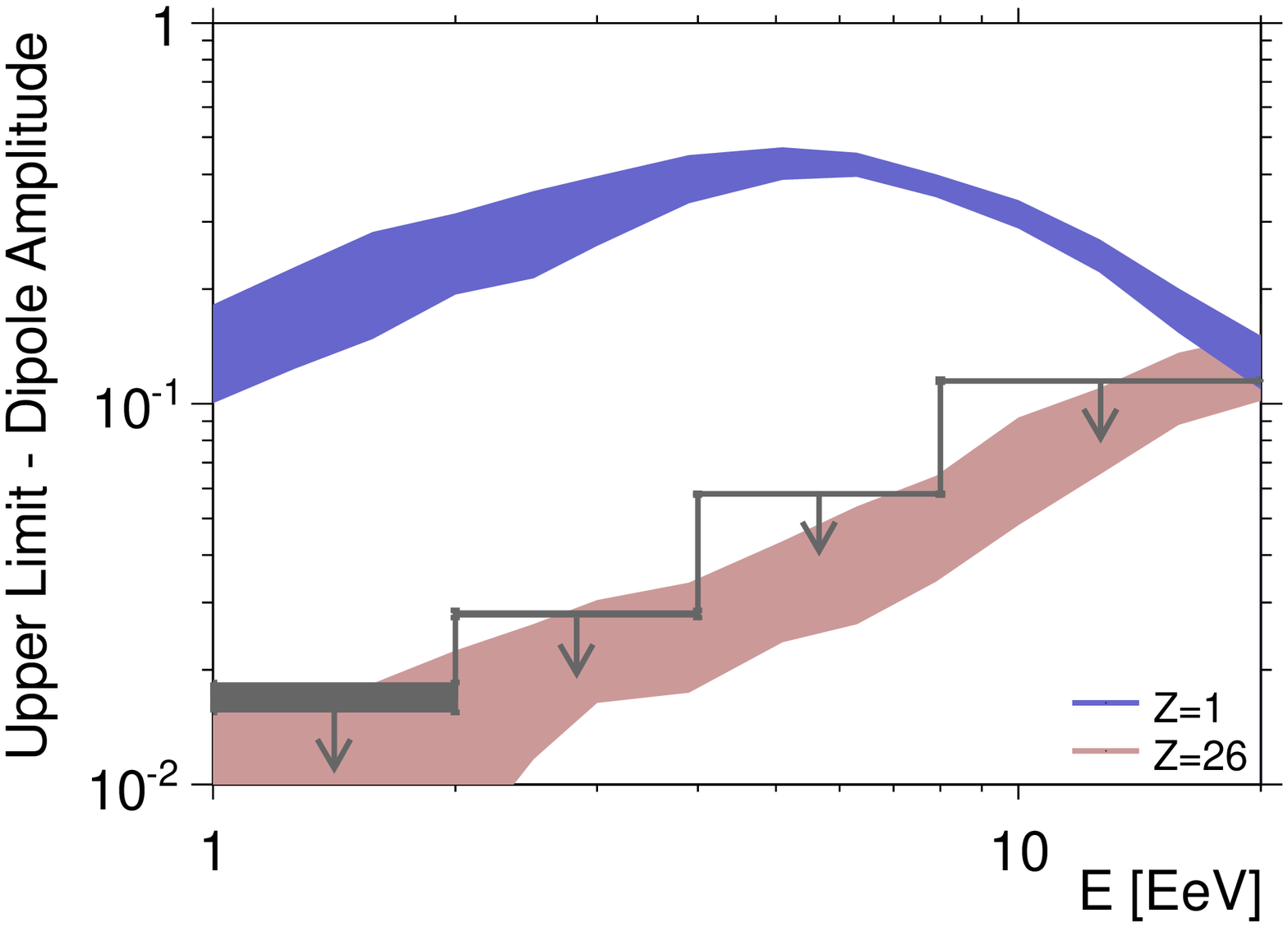}
  \includegraphics[width=7.5cm]{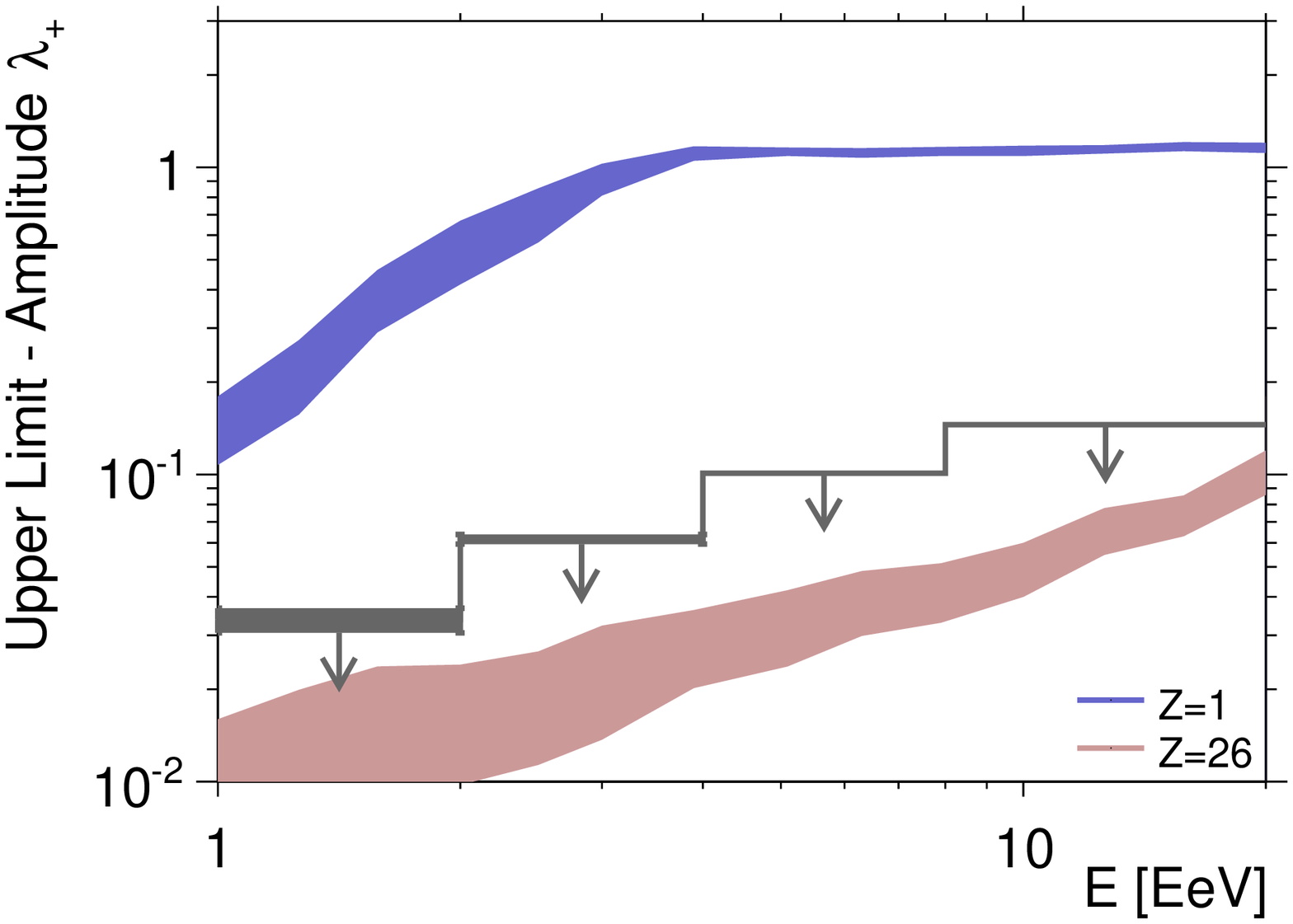}
  \caption{99\% $C.L.$ upper limits on dipole and quadrupole amplitudes as a function of the energy.
  Some generic anisotropy expectations from stationary Galactic sources distributed in the disk are also shown, 
  for protons and iron-nuclei. The fluctuations of the amplitudes due to the stochastic nature of the turbulent 
  component of the magnetic field are sampled from different simulation data sets and are shown by the bands 
  (see text).}
\label{fig:UL}
\end{figure}

From these analyses, the upper limits obtained at 99\% confidence level on dipole and quadrupole amplitudes 
are shown in figure~\ref{fig:UL}. These upper limits are stringent enough to challenge some Galactic scenarios.
One of them, in which sources of EeV-cosmic rays are stationary, densely and uniformly distributed in the 
Galactic disk, and emit particles in all directions, is illustrated in the following.

Both the strength and the structure of the magnetic field in the Galaxy, known only approximately, are of 
central importance for the propagation of EeV-cosmic rays. While the turbulent component dominates in 
strength by a factor of a few, the regular component, of the order of few microgauss in the disk, is expected 
to imprint dominant drift motions as soon as the Larmor radius of cosmic rays is larger than the maximal 
scale of the turbulences - thought to be in the range 10-100~pc. The parameterisation adopted here is the
one obtained recently in~\cite{Pshirkov2011}. Note however that estimations based on a much improved
model reported in~\cite{Jansson2012} are currently being carried out~\cite{Awal2014}. Preliminary results
are overall in good agreement in terms of the expected anisotropies with the ones reported 
in~\cite{AugerApJL2013,AugerApJS2012} and presented here.

To describe the propagation of EeV-cosmic rays in such a magnetic field, the direct integration of trajectories 
is the most appropriate tool. To obtain the anisotropy of cosmic rays emitted from sources uniformly distributed 
in a cylinder with a radius of 20~kpc from the galactic centre and with a height of $\pm$ 100~pc, the widely-used 
method first proposed in~\cite{Thielheim1968} and adopted in~\cite{AugerApJL2013,AugerApJS2012} consists 
in back-tracking anti-particles with random directions from the Earth to outside the Galaxy. Each test particle 
\textit{probes} the total luminosity along the path of propagation from each direction as seen from the Earth. 
For \textit{stationary sources emitting cosmic rays in all directions}, the expected flux in the initial sampled 
direction is proportional to the time spent by each test particle in the source region.

The dipole and quadrupole amplitudes obtained for several energy values covering the range 
$1\leq E/\mathrm{EeV}\leq 20$ are shown in figure~\ref{fig:UL}. Simulated event sets with $5~10^5$ test 
particles were generated to shrink the statistical uncertainties on amplitudes at the $0.5$\% level - which is
necessary to probe unambiguously amplitudes down to the percent level. The shaded regions stand for the
RMS of the amplitudes sampled from 20 independent sets of $5~10^5$ test particles, where the stochastic
component of the magnetic field (\textit{i.e.} the turbulence) is frozen differently in each set. Clearly, 
the resulting amplitudes for protons stand well above the allowed limits. Consequently, unless the
strength of the magnetic field is much higher than in the picture used here, the upper limits derived
in this analysis exclude the light component of cosmic rays coming from galactic stationary sources
densely distributed in the galactic disk and emitting in all directions. To respect the dipole limits below the 
ankle energy, the fraction of protons should not exceed $\simeq$ 10\% of the cosmic-ray composition
unless these protons do not originate from sources that can be roughly characterised by the picture used 
here. This is particularly interesting in the view of the indications for the presence of a light component
around $1~$EeV from shower depth of maximum measurements~\cite{AugerPRL2010,HiResPRL2010,
TAAPS2011,Yakutsk2012}, though firm interpretations of these measurements in terms of the atomic 
mass still suffer from some ambiguity due to the uncertain hadronic interaction models used to describe 
the shower developments.

\section{Large-Scale Structure with Full-Sky Coverage Above $10~$EeV}
\label{sec:fullsky}

As stressed in section~\ref{sec:spherical_harmonic}, the only way to unambiguously measure  
multipole moments to any order requires full-sky coverage. At present, full-sky coverage can only be achieved
through the meta-analysis of data recorded at observatories located in both hemispheres of the Earth.
The Telescope Array is the largest experiment ever built in the Northern hemisphere to study ultra-high energy
cosmic rays. Given the respective latitudes and zenith ranges covered by both experiments, full-sky coverage
can be achieved by combining both Pierre Auger Observatory and Telescope Array data sets. To facilitate the first 
joint analysis of this kind, the energy threshold, 10~EeV, is chosen to guarantee that both observatories operate 
with full-detection efficiency~\cite{Deligny2013,AugerTA2014}. In this way, the respective exposure functions follow 
purely geometric expectations.

The main challenge in combining the data sets is to account adequately for the relative exposures of both
experiments. In principle, the combined directional exposure of the two experiments should be simply the sum 
of the individual ones. However, individual exposures have here to be re-weighted by some empirical factor $b$ 
due to the unavoidable uncertainty in the relative exposures of the experiments:
\begin{equation}
\label{eqn:omega}
\omega(\mathbf{n};b)=\omega_{\mathrm{TA}}(\mathbf{n})+b\omega_{\mathrm{Auger}}(\mathbf{n}).
\end{equation}
Written in this way, $b$ is a dimensionless parameter of order unity. In practice, only an estimation $\overline{b}$ 
of the factor $b$ can be obtained, so that only an estimation of the directional exposure 
$\overline{\omega}(\mathbf{n})\equiv\omega(\mathbf{n};\overline{b})$ can be achieved through 
equation~\ref{eqn:omega}. The parameter $b$ can thus be viewed as a fudge factor which absorbs any kind 
of systematic uncertainties in the relative exposures, whatever the sources of these uncertainties - resulting from
different energy scales for instance. This empirical factor is arbitrary chosen in the following to re-weight the 
directional exposure of the Pierre Auger Observatory relatively to the one of the Telescope Array.

A band of declinations around the equatorial plane is exposed to the fields of view of both experiments,
namely for declinations between $-15^\circ$ and $25^\circ$. This overlapping region can be used for 
designing an empirical procedure to get a relevant estimate of the parameter $b$~\cite{Deligny2013}.
Considering as a first approximation the flux $\Phi(\mathbf{n})$ as isotropic, the overlapping region 
$\Delta\Omega$ can be utilised to derive a first estimate $\overline{b}^{(0)}$ of the $b$ factor by 
\textit{forcing the intensities of both experiments to be identical in this particular region}. This can be 
easily achieved in practice, by taking the ratio of the $\Delta N_{\mathrm{TA}}$ and $\Delta N_{\mathrm{Auger}}$ 
events observed in the overlapping region $\Delta\Omega$ weighted by the ratio of nominal exposures:
\begin{equation}
\label{eqn:b0}
\overline{b}^ {(0)}=\frac{\Delta N_{\mathrm{Auger}}}{\Delta N_{\mathrm{TA}}}\frac{\displaystyle\int_{\Delta\Omega}d\Omega~\omega_{\mathrm{TA}}(\mathbf{n})}{\displaystyle\int_{\Delta\Omega}d\Omega~\omega_{\mathrm{Auger}}(\mathbf{n})}.
\end{equation}
Then, inserting $\overline{b}^ {(0)}$ into $\overline{\omega}$, 'zero-order' $\overline{a}_{\ell m}^{(0)}$ 
coefficients can be obtained. This set of coefficients is only a rough estimation, due to the limiting assumption 
on the flux (isotropy).

On the other hand, the expected number of events in the common band for each observatory,
$\Delta n_{\mathrm{TA}}^{\mathrm{exp}}$ and $\Delta n_{\mathrm{Auger}}^{\mathrm{exp}}$, can 
be expressed from the underlying flux $\Phi(\mathbf{n})$ and the true value of $b$ as:
\begin{eqnarray}
\label{eqn:dn}
\Delta n_{\mathrm{TA}}^{\mathrm{exp}}&=&\int_{\Delta\Omega}d\Omega~\Phi(\mathbf{n})\omega_{\mathrm{TA}}(\mathbf{n}) \nonumber \\
\Delta n_{\mathrm{Auger}}^{\mathrm{exp}}&=&b\int_{\Delta\Omega}d\Omega~\Phi(\mathbf{n})\omega_{\mathrm{Auger}}(\mathbf{n}).
\end{eqnarray}
From equations~\ref{eqn:dn}, and from the set of $\overline{a}_{\ell m}^{(0)}$ coefficients, an iterative 
procedure estimating at the same time $b$ and the set of $a_{\ell m}$ coefficients can be considered in the 
following way: 
\begin{equation}
\label{eqn:b_est}
\overline{b}^{(k+1)}=\frac{\Delta N_{\mathrm{Auger}}}{\Delta N_{\mathrm{TA}}}\frac{\displaystyle\int_{\Delta\Omega}d\Omega~\overline{\Phi}^{(k)}(\mathbf{n})\omega_{\mathrm{TA}}(\mathbf{n})}{\displaystyle\int_{\Delta\Omega}d\Omega~\overline{\Phi}^{(k)}(\mathbf{n})\omega_{\mathrm{Auger}}(\mathbf{n})},
\end{equation}
where $\Delta N_{\mathrm{TA}}$ and $\Delta N_{\mathrm{Auger}}$ as derived in the first step are used 
to estimate $\Delta n_{\mathrm{TA}}^{\mathrm{exp}}$ and $\Delta n_{\mathrm{Auger}}^{\mathrm{exp}}$ 
respectively, and $\overline{\Phi}^{(k)}$ is the flux estimated with the set of $\overline{a}_{\ell m}^{(k)}$ 
coefficients. 

This procedure makes it possible to use the powerful multipolar analysis method for characterising the sky map of 
ultra-high energy cosmic rays. It pertains to any full-sky coverage achieved by combining data sets from different 
observatories, and opens a rich field of anisotropy studies. This technique will be applied to data sets from the 
Telescope Array and the Pierre Auger Observatory and will be reported in a near future~\cite{AugerTA2014}.

\section{Outlook}
\label{sec:outlook}

A thorough search for large-scale anisotropies in the distribution of arrival directions of cosmic rays detected above
$\simeq 10$~PeV at the Pierre Auger Observatory has been presented. With respect to the traditional search in right 
ascension only, spherical harmonic analyses has been performed above 1~EeV to characterise the distribution of arrival
directions in terms of both the right ascension and the declination. No significant deviation from isotropy is revealed within 
the systematic uncertainties, although there are few hints that may be indicative of a genuine signal whose amplitude is at 
the level of the current statistical noise. The current sensitivity allows us to challenge an origin of cosmic rays from stationary 
galactic sources densely distributed in the galactic disk and emitting predominantly light particles in all directions.

Future work will profit from the increased statistics. Also, future joint studies between the Telescope Array and
the Pierre Auger Collaborations will allow a determination of the full set of spherical harmonic coefficients at ultra-high 
energies. This will provide further constraints helping to understand the origin of cosmic rays above $\simeq$ 10~PeV.

\section*{Acknowledgements}
The author thanks the organisers of the workshop for the opportunity to present these results and the members of the Pierre
Auger Collaboration for all the work described in this contribution.

\end{document}